\documentclass[twocolumn,showpacs,preprintnumbers,amsmath,amssymb,prl,superscriptaddress,nofootinbib]{revtex4}
\usepackage{graphicx} 
\usepackage{dcolumn} 
\usepackage{bm} 
\usepackage{amssymb} 
\usepackage{amsmath} 
\usepackage{color}

\definecolor{darkgreen}{RGB}{50,150,0}

\begin{document}

\title{Distance and de Sitter Conjectures on the Swampland}

\author{Hirosi Ooguri}
\affiliation{Walter Burke Institute for Theoretical Physics, Caltech, Pasadena, CA 91125, USA}
\affiliation{Kavli Institute for the Physics and Mathematics of the Universe, University of Tokyo, Kashiwa 277-8583, Japan}
\author{Eran Palti}
\affiliation{Max-Planck-Institut f\"ur Physik (Werner-Heisenberg-Institut), 
Fohringer Ring 6, 80805 Munchen, Germany }
\author{Gary Shiu}
\affiliation{Department of Physics, University of Wisconsin-Madison, Madison, WI 53706, USA}
\author{Cumrun Vafa}
\affiliation{Jefferson Physical Laboratory, Harvard University, Cambridge, MA 02138, USA}

\begin{abstract}
Among Swampland conditions, the distance conjecture characterizes the
geometry of scalar fields and the de Sitter conjecture
constrains allowed potentials on it.  We point out a connection between the distance conjecture 
and a refined version of the de Sitter conjecture  in any parametrically controlled regime
of string theory by using Bousso's covariant entropy bound.  The refined version turns out to 
evade all counter-examples at scalar potential maxima that have been raised.
We comment on the relation of our result to the Dine-Seiberg problem.

\end{abstract}

\preprint{CALT-TH 2018-042, IPMU 18-0164, MPP-2018-248, MAD-TH-18-06}
\maketitle

\section{Introduction}
\label{sec:int}

Recently, motivated by a number of string theoretical constructions with controlled approximations, 
it was proposed in \cite{Obied:2018sgi} that the potential for scalar fields in 
string theory satisfies the universal bound, 
\begin{equation}
\left| \nabla \ V\right| \geq \frac{c}{M_p} \cdot V \;,
\label{dsc}
\end{equation}
for some constant $c>0$ of order $1$, where $M_p$ is the Planck mass.\footnote{The power of $M_p$ in the conjecture depends on the dimensions. In this paper, for brevity, we write the formulas for the specific case of $4$ dimensions.} Cosmological implications of 
this conjecture were studied in \cite{Agrawal:2018own}.  The bound is called the de Sitter
conjecture as it excludes (meta-)stable de Sitter vacua in string theory.
There have been a number of follow up papers \cite{Andriot:2018wzk,Dvali:2018fqu,Achucarro:2018vey,Garg:2018reu,Lehners:2018vgi,Kehagias:2018uem,Dias:2018ngv,Denef:2018etk,Colgain:2018wgk,Roupec:2018mbn,Andriot:2018ept,Matsui:2018bsy,Ben-Dayan:2018mhe, Damian:2018tlf,Conlon:2018eyr,Kinney:2018nny,Dasgupta:2018rtp,Cicoli:2018kdo, Kachru:2018aqn, Akrami:2018ylq, 
Murayama:2018lie, Marsh:2018kub, Brahma:2018hrd, Choi:2018rze, Das:2018hqy, Danielsson:2018qpa, 
Wang:2018duq, Han:2018yrk, Visinelli:2018utg, Moritz:2018ani, Bena:2018fqc, 
 Brandenberger:2018xnf,  Brandenberger:2018wbg, Quintin:2018loc, Heisenberg:2018rdu, Gu:2018akj,
Heisenberg:2018yae,  Brandenberger:2018fdd,Ashoorioon:2018sqb,Odintsov:2018zai,Motaharfar:2018zyb,Kawasaki:2018daf,Hamaguchi:2018vtv,Lin:2018kjm,Ellis:2018xdr,Dimopoulos:2018upl}.\footnote{For alternative perspectives on cosmology and microscopic aspects of de Sitter see \cite{Banks:2018ypk},  \cite{Dvali:2017eba} and \cite{Conlon:2012tz}.}   
The main aim of this paper is to connect this conjecture to another better established swampland condition,
which is known as the distance conjecture \cite{Ooguri:2006in}. We will show that,
 in any weak coupling regime of string theory,\footnote{In this paper, 
weak coupling refers to any limit in any direction in the space of low energy scalar fields where
a parametrically controlled approximation to a physical observable is possible, 
while the weak string coupling refers to the specific limit of the dilaton field.} a refined version of this conjecture  follows 
 from the distance conjecture, combined with Bousso's covariant entropy bound \cite{Bousso:1999xy} applied to an accelerating universe. 
 
The refined version of the de Sitter conjecture which we propose in this paper is stated as follows:
 
 \medskip
 \noindent
 {\bf Refined de Sitter Conjecture.} 
{\it A potential $V(\phi)$ for scalar fields in a low energy 
effective theory of any consistent quantum gravity must satisfy either,
\begin{equation}
\left|\nabla V\right| \geq \frac{c}{M_p} \cdot V \;,
\label{rdscone}
\end{equation}
or 
\begin{equation}
{\rm min} \left( \nabla_i \nabla_j V \right) \leq - \frac{c'}{M_p^2} \cdot V \;,
\label{rdsctwo}
\end{equation}
for some universal constants $c, c' >0$ of order $1$,
 where the left-hand side of (\ref{rdsctwo}) is the minimum eigenvalue of the Hessian $ \nabla_i \nabla_j V$ in an orthonormal frame. }

\medskip
Note that the conjecture is trivial if $V$ is non-positive because (\ref{rdscone}) is satisfied, or in the limit $M_p \rightarrow \infty$, where gravity
decouples. This refined version still excludes (meta-)stable de Sitter vacua.

In this paper we provide evidence for the refined conjecture only in parametrically controlled regimes of string theory, but it is natural to conjecture that it holds more generally.
Possible refinements of the original de Sitter conjecture have been considered in \cite{Dvali:2018fqu,Andriot:2018wzk,Garg:2018reu}.  The refinement above is in essence the same as the proposal of \cite{Garg:2018reu}.\footnote{
The refined conjecture stated in \cite{Garg:2018reu} reads stronger than the one
we propose here since it was stated with the absolute-value sign on
$\eta_V \sim \nabla^2 V/V$, $i.e.$ the precise complement of the
slow roll conditions. However, we have been informed by the authors of the paper that this was unintentional as they were motivated by avoiding application of the dS conjecture to maxima in potentials.}
Note that the refined version is stronger than the complement of the slow roll conditions for the
cosmic inflation.  

Effective descriptions of string theory are controlled by a multitude of coupling constants, such as the string coupling or the volume of extra dimensions. All such couplings are field-dependent, and in particular, any weak couplings are associated with large distances in field space. 
The distance conjecture \cite{Ooguri:2006in} is about a
connected component of the moduli space of string vacua,
and it states that 
it is always possible for a scalar field to change its value from its original point with an arbitrarily large amount of geodesic distance and that
towers of light states with masses, 
\begin{equation}
\label{ma}
m \sim e^{-a \Delta \phi} \;,
\end{equation}
descend from the ultraviolet if the change $\Delta \phi$ is trans-Planckian.\footnote{Note that the light states do not necessarily have to be particles but can be extended objects. In this paper, we do not
refer to this distinction explicitly.}
This conjecture has been tested to significant extent in string theory
 \cite{Cecotti:2015wqa,Palti:2015xra,Baume:2016psm,Klaewer:2016kiy,Valenzuela:2016yny,Blumenhagen:2017cxt,Palti:2017elp,Lust:2017wrl,Hebecker:2017lxm,Cicoli:2018tcq,Grimm:2018ohb,Heidenreich:2018kpg,Blumenhagen:2018nts,Lee:2018urn,Grimmtoappear}, and has been generalized to cases with non-trivial potentials $V(\phi)$ \cite{Baume:2016psm,Klaewer:2016kiy}. 

It is reasonable to expect that the number of states in the Hilbert space of the low energy
theory increases monotonically when the towers of light particles emerge as predicted by the distance conjecture. 
On the other hand, to the observable part
of an accelerating universe, 
we can naturally associate an entropy, which is finite and is set by the value of 
the scalar potential.
For a de Sitter vacuum,
the Gibbons-Hawking entropy \cite{PhysRevD.15.2738} has been
interpreted as the logarithm of the number of allowed states 
in the causal domain. 
We will use the Bousso bound to extend it to a more general class of accelerated universes. Relating the two notions of entropy will lead us to the refined de Sitter conjecture.

A monotonic behavior of the scalar potential
was shown by Dine and Seiberg \cite{Dine:1985he} in the weak string coupling limit.
 Their argument is that, if the potential is generated, its expected dependence 
 on the string coupling constant implies that the leading correction will always 
 generate a rolling potential with non-zero
gradient. Our result generalizes it to any direction in the space of all possible couplings in string theory,
not limited to the string coupling constant and 
whether the corrections arise perturbatively or non-perturbatively.

After the original de Sitter conjecture was proposed \cite{Obied:2018sgi},
possible counter-examples have been suggested \cite{Denef:2018etk, Conlon:2018eyr, Murayama:2018lie,Choi:2018rze,Hamaguchi:2018vtv}. We will show that the
refined conjecture evades them all. We also note that it is compatible with the observation that tachyons are ubiquitous in classical dS and quintessence solutions
\cite{Flauger:2008ad,Caviezel:2008tf,Danielsson:2009ff,Danielsson:2011au,Shiu:2011zt,VanRiet:2011yc,Danielsson:2012et,Blaback:2013fca,Junghans:2016uvg,Junghans:2016abx,Danielsson:2018ztv}.
Our derivation of the refined conjecture in controllable
 regimes of string theory
suggests that, to establish the conjecture more generally or 
to find any counter-examples, one must consider the regime where coupling
constants are not parametrically small, requiring more advances in string theory techniques.

\section{Distance Conjecture}
\label{sec:sdc}

The distance conjecture \cite{Ooguri:2006in} states the following: Consider the moduli space of string vacua parameterized by scalar fields $\phi^i$ where $V\left(\phi^i\right)=0$, with the metric given by their kinetic terms. 
If we start from any point in the moduli space and go to a different point at a large geodesic distance away $\Delta \phi \gg 1$,  towers of particles with masses $m\sim e^{-a\Delta \phi }$ descend from the ultraviolet with some ${\cal O}(1)$ constant $a >0$. The conjecture has been extensively tested in string theory, and there are results establishing its validity in general settings \cite{Cecotti:2015wqa,Grimm:2018ohb,Lee:2018urn,Grimmtoappear}.
The critical value of $\Delta \phi$ at the onset of the exponential
behaviour has been studied and quantified in
\cite{Baume:2016psm,Klaewer:2016kiy,Blumenhagen:2017cxt,Hebecker:2017lxm,Cicoli:2018tcq,Blumenhagen:2018nts,Landete:2018kqf}.
 Here and in the following, we work in Planck units, where $M_p=1$.

The distance conjecture was motivated by the idea that, as we go over infinite distances in moduli space, the original effective theory breaks down and 
a dual weakly coupled description takes over, where the basic light degrees of freedom are dictated by the towers.  The ideas of how various aspects of physics emerge from such a dual description have also been studied  in \cite{Harlow:2015lma,Heidenreich:2017sim,Heidenreich:2018kpg,Grimm:2018ohb} and related to the weak gravity conjecture \cite{ArkaniHamed:2006dz}. In the following, we will assume this interpretation of the distance conjecture in terms of a duality.

A generalization to the case with a non-trivial potential $V\left(\phi^i\right)$ has been conjectured \cite{Baume:2016psm,Klaewer:2016kiy}. It has also been tested in string theory, in particular with axions, which have offered the strongest challenges to it so far. 
Naively, the axionic shift symmetry prevents an axion-dependent field space metric. However, backreaction on the saxion partner can lead to the same asymptotic behavior, see \cite{Blumenhagen:2018hsh} for a review. 
To generalize the distance conjecture, consider a region in the space of scalar fields where $0\leq V\left(\phi^i\right )< {\cal O}(1) $. 
To have parametrically controlled weak coupling points, the region must have an infinite diameter,
$i.e.$, one should be able to go infinite  distance away from a given point.  
A natural extension of the distance conjecture  would be that, for sufficiently far distances, we have towers of light particles whose masses go as $m \sim e^{-a\Delta \phi}$, just as in the moduli space case.\footnote{Though the definition of distance needs to be amended when a potential is present \cite{Landete:2018kqf}, this 
subtlety is not relevant when we discuss parametrically large distance.}

When $V(\phi)=0$, we have evidences for the dual descriptions with light states. In this note, we assume 
that such descriptions also apply when $V\left(\phi\right)\neq 0$.
For example, in ${\cal N}=2$ supergravity, the only way to induce a potential is by gauging isometries in field space and the potential is tied to the gauge couplings of the gauge fields. These have been shown to be recovered from tower of light particles \cite{Grimm:2018ohb}.  

\section{Horizon and Bousso Bound} 
\label{sec:boussobound}

The Gibbons-Hawking entropy $S_{{\rm GH}}$ \cite{PhysRevD.15.2738} of a de Sitter space is proportional to the area of its event horizon,
\begin{equation}
\label{GH}
S_{{\rm GH}}=R^2 = 1/\Lambda \;,  
\end{equation}
where $R$ is the curvature radius and  $\Lambda$ is the cosmological constant,
neglecting numerical factors of ${\cal O}(1)$.
This quantity has been interpreted in term of the dimension of the Hilbert space ${\cal H}$
in an observer's causal domain \cite{Banks:2000fe,Witten:2001kn},
\begin{equation}
\mathrm{dim\;}{\cal H} = e^{1/\Lambda} \;.
\end{equation}
In the limit of $\Lambda \rightarrow 0$, the dimension becomes infinite as expected
for the Minkowski space. 

Suppose there is a positive potential $V(\phi)$
 with non-trivial dependence on $\phi$.  If $V$ has a local minimum and the resulting meta-stable de Sitter
 space is long lived, its entropy is meaningful. Even if the potential has non-zero gradient, 
as far as $|\nabla V | /V$ is  less than $\sqrt{2}$, we have a universe with accelerated 
expansion with an apparent horizon at $R = 1/\sqrt{V}$. Since the apparent horizon 
is always inside of a cosmic event horizon if the latter exists,
lightsheets emanating from it will close at caustics, enabling us to use the Bousso bound \cite{Bousso:1999xy} for 
the entropy in the portion of the Cauchy surface enclosed by the apparent horizon.

This semi-classical picture is valid provided quantum fluctuations of $\phi$ are negligible. 
If the Hessian $\nabla_i \nabla_j V$ has a negative eigenvalue  below $ - c'/R^2$, with $c'$ of $O(1)$, 
zero-point fluctuations at the 
horizon crossing becomes tachyonic and the semi-classical picture breaks down. 

We conclude that, if $V$ is positive and satisfies,
\begin{equation}
|\nabla V|  \leq {\sqrt 2}\cdot V ~~ {\rm and} ~~ {\rm min}(\nabla_i \nabla_j V) \geq - c' V
 \;,
\end{equation}
there is an accelerating universe, 
and the entropy inside of its apparent horizon is bounded by $R^2$.
Note that, though the second inequality is required for the stability of zero point fluctuations at the 
horizon crossing, it also ensures that the first inequality
holds within one Hubble time.

\section{Entropy and Towers of Particles} 
\label{sec:dse}

In a weak coupling regime, the distance conjecture claims that the number of effective degrees
of freedom increases by having towers of light particles with exponentially small masses.
This should increase the entropy and influence how the potential behaves in any weak
coupling limit. 
To quantify this, let us parametrize the number of particle species below a certain
cutoff of the effective theory as, 
\begin{equation}
\label{largeN}
N(\phi) \sim n(\phi) e^{b\phi } \;.
\end{equation} 
Here $n\left(\phi\right)$ is effective the number of towers of states that are becoming light,
which we expect to increase monotonically $dn/d\phi\geq 0$ toward the weak coupling limit. 
The exponent $b$ depends on mass gaps and other features of the towers and
is in general different from  $a$ in (\ref{ma}). 

Let $R$ denote the radius of the apparent horizon, as in the last section.
We expect that the entropy $S_{\rm tower}(N, R)$ coming from
the towers of particles increases as the number $N$ of the species increases.
For large $N$ and $R$, we parametrize the $N$ and $R$ dependence of the entropy as,
\begin{equation}
\label{senpar} 
  S_{\rm tower} (N, R) \sim N^\gamma R^\delta \; , 
\end{equation} 
with some positive exponents $\gamma$ and $\delta$. 
When the universe is accelerating and an apparent horizon forms, the Bousso
bound applied to $S_{\rm tower}(N, R)$ gives,
\begin{equation}
\label{entropybound}
    N^\gamma R^\delta \leq R^2 \; .
\end{equation}
As we take the weak coupling limit, $N$ increases exponentially as (\ref{largeN}),
and $R$ should also change accordingly so that the bound is not violated.\footnote{The Bousso bound 
generalizes the Bekenstein bound \cite{Bousso:2002bh}, which sets an 
upper bound on 
the number of allowed states in a box with a given energy 
\cite{Bekenstein:1974ax,Bekenstein:1980jp}. 
 The Bekenstein bound is satisfied with appropriate definitions of 
the energy and the entropy, 
and becomes saturated in the large $N$ limit \cite{Page:1982fj,Unruh:1982ic,Pelath:1999xt,Marolf:2003wu, Marolf:2003sq,Casini:2008cr}. Similarly, we expect that the 
Bousso bound be saturated
in the large $N$ limit.}   Since light degrees of freedom dominate the Hilbert space in the weak coupling regime,
we expect them to saturate the Bousso bound leading to,
\begin{equation}
\label{potential}
   V(\phi) \sim R^{-2} \sim N^{- \frac{2 \gamma}{2-\delta}} \; .
\end{equation}
The exponential behavior (\ref{largeN}) of the number $N$ of species
combined with the inequality $dn/d\phi \geq 0$ implies  the first condition
 (\ref{rdscone}) of the refined de Sitter conjecture with,\footnote{Note that $\delta < 2$ since we assume the light tower of states dominates low energy states in the Hilbert space. }
\begin{equation}
 \label{cvalue}
    c = \frac{2b \gamma}{2-\delta} \; .
\end{equation}

As we discussed in the previous section, 
the second condition (\ref{rdsctwo}) is prerequisite for the apparent horizon to exist.
We note that the condition also evades the proposed counter-examples to the original de Sitter conjecture \cite{Denef:2018etk, Conlon:2018eyr, Murayama:2018lie, Choi:2018rze, Hamaguchi:2018vtv}
since these are all based on maxima of potentials and can be checked to satisfy (\ref{rdsctwo}). More generally, we note that the maxima of any axion-type potentials,
where $V\sim\Lambda^4 {\rm cos}(\phi/f)$, satisfy,
\begin{equation}
\frac{{\rm min} \left( \nabla_i \nabla_j V \right)}{V} \lesssim - \frac{1}{f^2} \;,
\label{axmax}
\end{equation}
where $f$ is the axion decay constant.  The weak gravity conjecture \cite{ArkaniHamed:2006dz} states that $f \lesssim 1$ for any controlled axion potential, in particular also for the QCD axion, and so condition (\ref{rdsctwo}) is satisfied.\footnote{While the effective axion decay constant can be enhanced with mechanisms such as alignment
\cite{Kim:2004rp}, and with the weak gravity conjecture satisfied by additional spectator instantons 
\cite{Rudelius:2015xta,Brown:2015iha, Brown:2015lia}, these loopholes are not of concern for the QCD axion.} For non-axion scalar fields, we also note that the $\eta$-problem \cite{PhysRevD.49.6410} implies that any maximum of an $F$-term based potential will generically satisfy (\ref{rdsctwo}). A violation would require some fine-tuned cancellations. 

\section{Entropy of free particles}
\label{sec:qftc}

Though there is no known method to compute $S_{\rm tower}(N, R)$ by enumerating
all states in the Hilbert space of quantum gravity in an accelerating universe, we can count their subset
when the low energy theory consists of $N$
free particles. To justify neglecting gravitational interactions between the particles, we
require that the system does not collapse into a black hole.
 We may regard the resulting entropy as a lower bound on $S_{\rm tower}$. 
Our counting of microstates essentially follows \cite{Page:1981an,Cohen:1998zx}, which was
used in the cosmological context in \cite{Banks:2005bm}.

Let us start with counting the number of states of a single free field with mass $m$ in 
a box of size $R$, up to a maximum momentum $k_{\rm max}$. Since
particle momenta are quantized in the unit of $1/R$, the entropy and energy
associated to the field are, 
\begin{equation}
\label{singleparticle}
S_{N=1} \sim  \left(k_{\rm max} R\right)^3 \;, \;\; 
E_{N=1} \sim \omega \left(k_{\rm max} R\right)^3 \;,
\end{equation}
where $\omega^2 = k^2 + m^2$. 
Let us assume for now that $k_{\rm max} \gg m$. 
In order for such a configuration not to collapse into a black hole, we require that the
Schwarzschild radius for the energy $E_{N=1}$ is less that the box size, namely
$E_{N=1} \lesssim  R$. Thus, the largest value $k_{\rm max}$ can take is $k_{\rm max} \sim R^{-1/2}$ and  
correspondingly 
$S_{N=1} \sim R^{3/2}$. 
Though this could never account for the Bousso bound, which scales as $R^2$,
it may be possible with large $N$ species of particles.

Indeed, it is a straightforward exercise of thermodynamics to generalize the above to $N$ 
species of particles. To maximize the entropy, we can regard each species to be in 
a thermal bath of temperature $T$, common to all the species. The total 
entropy and energy for particles with masses less than $T$ are then,
\begin{equation}
S_N \sim N T^3 R^3 \;, \;\; E_N \sim N T^4 R^3 \; .
\end{equation}
Particles with masses
grater than $T$ do not contribute significantly to the entropy and
can be safely ignored. 
Requiring that the system does not collapse to a black hole, the maximum energy
one can consider is at $E_N \sim R$ and therefore the entropy of the largest subspace 
we can consider while ignoring gravitational effects is,
\begin{equation}
\label{freeentropy}
S_N \sim N^{1/4} R^{3/2} \; .
\end{equation}
Note that this is $N^{1/4}$ times the result (\ref{singleparticle}) for a single species, 
and $\gamma = 1/4$ and $\delta = 3/2$ in the parametrization of (\ref{senpar}).
In the large $N$ limit, $S_N$ can saturate the Bousso bound
$R^2$. Using (\ref{cvalue}), we find $c = b$.  
This, however, requires an extremely large number of
species, $N \sim R^2$, with the minimum entropy assigned to each.

Since
(\ref{freeentropy}) is meant to be the lower bound on $S_{\rm tower}(N, R)$, 
the exponents in (\ref{senpar}) 
should be bounded as $\gamma \geq 1/4$ and $\delta \geq 3/2$.
Therefore,  
\begin{equation}
\frac{2-\delta}{\gamma} \leq 2.
\end{equation}
Since (\ref{potential}) relates $N \sim R^{\frac{2-\delta}{\gamma}}$, the
number $N$ species needed to saturate the Bousso bound should scale slower than $R^2$ in general.

\section{Cosmological Implications}
\label{sec:cos}

While the de Sitter conjecture is not sensitive to the particular ${\cal O}(1)$
values of $\delta$ and $\gamma$, the phenomenology is. To give a brief
overview of this dependence, we consider a tower which is evenly spaced, so
the mass scale of the $n^{\mathrm{th}}$ states is $m_n \sim n m$. For the
cutoff scale $\Lambda_N$, below which there are $N$ states contributing to
the entropy, we consider the range $N^{-\frac12} < \Lambda_N < 1$. We
therefore have states
with a mass scale in the range
\begin{equation}
R^{\frac{3\left(\delta-2\right)}{2\gamma}} < m <
R^{\frac{\delta-2}{\gamma}}\;.
\end{equation}

If our universe is in a weakly coupled regime,
it would necessarily imply that the dark sector involves a tower of light states 
(see \cite{Dienes:2011ja} for similar scenarios).
Since the Hubble scale of our current universe is $R \sim 10^{60}$,
taking the free particle  
values of $\delta=\frac32$ and $\gamma=\frac14$ 
would give a phenomenologically unrealistic scenario.
Thus, if the lower bound is saturated,
our current universe would be in a strongly coupled regime.
Taking different values, for example $\delta = \frac74$ and
$\gamma=1$, would give $N\sim 10^{15}$ and $\mathrm{MeV} < m < \mathrm{TeV}$. There would be time dependence for the mass of such a tower as the quintessence field evolves (see in particular a recent study of this \cite{Matsui:2018xwa}).  The strong dependence of the
cosmology and phenomenology on the particular microstate counting scheme
of the tower of states leads to a close interaction between microscopic
physics and observations, and it would be interesting to develop it further.

\section{Relation to Weak Gravity Conjecture}

While the analysis in this paper has been focused on the weakly coupled region, it is natural
to speculate extension of the de Sitter conjecture to the entire parameter space of string theory. 
In this respect, it is interesting to point out a similarity between the de Sitter conjecture (\ref{dsc}) 
and the scalar weak gravity conjecture proposed in
\cite{Palti:2017elp}, which states that for each canonically normalized 
massless scalar field $\phi$, there must exist a particle whose mass $m(\phi)$ satisfies
$|\partial_{\phi} m| > m$. 

The connection between
the conjectures may arise by thinking about
objects whose mass behaves like the scalar potential $V\left(\phi\right)$.
For example, a space-filling three-brane would have a tension set by
$T=V\left(\phi\right)$, and an analogous
statement to the scalar weak gravity conjecture to the tension would lead to an equation 
of the form $|\nabla T|\gtrsim T$, which is of the type expected for the de Sitter conjecture. Furthermore, the second condition (\ref{rdsctwo}) could be interpreted in this context as the stability of objects predicted by the
weak gravity conjecture. We leave a more detailed study of such potential connections for future work.

\section{Connection to Dine-Seiberg Problem}

In \cite{Dine:1985he}, Dine and Seiberg argued that no non-supersymmetric stable or metastable vacuum of string theory can be found at a parametrically weak string coupling point.  Since no potential is generated when the string coupling constant $g$ vanishes,
if the potential is non-zero at small value of $g$, we can parametrize it as $V=g^n$, 
where a positive power $n$ depends on the first loop order with non-zero contributions to $V$.
Since $g$ is parametrized by the canonically normalized dilaton $\phi$ as $g=e^{-\phi}$, this give $V(\phi) \sim e^{-n \phi}$.  Note that this satisfies the de Sitter conjecture $|\nabla V|>c \cdot V$ by taking $0<c<n$.  It is possible that $V$ vanishes to all orders in string perturbation theory, but that some non-perturbative effect generates it.  In such a case, we can estimate 
$V\sim g^n e^{-A/g^k}$ and the de Sitter conjecture is
again satisfied, where 
$n$  is determined by the first loop order with non-trivial contributions to $V$
in the instanton background and 
$k$ is an even or odd positive integer depending on whether the contribution comes from closed or open string instantons. 
The conclusion 
of  \cite{Dine:1985he} is that either our universe is rolling toward the weak string coupling, because the leading correction to $V$ would have to suffice to give 
such a behavior at parametrically weak coupling,
 or it is stabilized at $g \sim {\cal O}(1)$, which will be hard to establish or rule out.
Our analysis has gone beyond their result in that, for the $V>0$ case, we found a universal behavior of the potential in any long distance directions in the scalar field space. 

With the relation discussed in this section in mind, it is important to make clear that our result is independent of the Dine-Seiberg argument, for which by-pass mechanisms have been proposed. We only rely on the assumptions that the distance conjecture holds and that we are in a sufficiently weakly-coupled regime that the light states dominate the Hilbert space. 
However, we would also like to emphasize that not having parametrically good control over couplings does not mean that no control is possible. Rather, we view our results as showing that establishing the validity or violation of the de Sitter conjecture will require careful studies of string theory vacua, quantifying corrections and
sharpening estimate of errors in our existing techniques and developing more powerful tools. Given the importance of the dark energy problem, effort into this direction is well justified.

\subsection{Acknowledgments}
We would like to thank Prateek Agrawal, Tom Banks, Raphael Bousso, Clifford Cheung, Daniel Chung, Joe Conlon, Albion Lawrence, Toshifumi Noumi, Georges Obied, Misao Sasaki, 
and Pablo Soler for discussions.
The work of GS is supported in part by 
the DOE grant DE-SC0017647 and the Kellett Award of the University of Wisconsin.
The work of CV is supported in part by NSF grant PHY-1067976.
The work of HO is supported in part by
U.S.\ Department of Energy grant DE-SC0011632,
by the World Premier International Research Center Initiative,
MEXT, Japan,
by JSPS Grant-in-Aid for Scientific Research C-26400240,
and by JSPS Grant-in-Aid for Scientific Research on Innovative Areas
15H05895.
HO also thanks the hospitality of
the Aspen Center for Physics, which is supported by
the National Science Foundation grant PHY-1607611. 
We would like to thank the hospitality of Simons Center for Geometry and Physics, where this work was initiated during the 2018 Simons Summer workshop.
\bibliography{ref}

\end{document}